\documentstyle[12pt]{article}
\begin{document}
\baselineskip .3in
\begin{titlepage}
\begin{center}
{\Large {\bf Fibonacci-Hubbard Chain at Zero and Finite Temperatures}}
\vskip .2in
{\em  Sanjay Gupta$^{\dag}$, Shreekantha Sil$^{\ddag}$ and 
Bibhas Bhattacharyya$^{\dag \dag}$}

\vskip .1in
$^{\dag}$TCMP Division, Saha Institute of Nuclear Physics,
1/AF-Bidhannagar, Calcutta -- 700064, India.\\
$^{\ddag}$Department of Physics, Vishwabharati, Shantiniketan, Birbhum\\
$^{\dag \dag}$Department of Physics, Scottish Church College,
1 \& 3, Urquhart Square, Calcutta -- 700006

\vskip .3in
{\bf Abstract}
\end{center}

\noindent
We have studied finite-sized single band Hubbard chains with Fibonacci 
modulation for half filling within a mean field approximation.
The ground state properties, together with the dc conductivity both at zero 
and non-zero temperatures, are calculated for such quasi-periodic Hubbard 
chains. While a reduction in the conductivity is found for strong electronic
interaction or strong Fibonacci modulation, a competition between these
two is observed to enhance the conductivity. 
The results at finite temperatures also illustrate some interesting 
features of such finite-sized systems. 

\vskip .1in
\noindent
{\bf PACS No.:} 71.10.Fd, 71.23.Ft, 75.30.Fd
\vskip .2in
\end{titlepage}
\newpage
\noindent
{\Large \bf I. Introduction}
\vskip .2in 
In recent years the study of the interplay of correlation and disorder in
condensed matter systems has become a subject of great interest.
Some recent experimental results, e.g. the metal-insulator transition
(MIT) in two dimensional electron gas in Si-MOSFETs \cite{Krav}, 
suggest that there exists a competition between the electronic 
correlation and the disorder in such systems.
Sheer complexity of the problem of considering  simultaneous effects
of both correlation and disorder  attracted the attention of the researchers
over the past few years. For example, the
effect of Hubbard correlation in presence of random disorder has been studied by
using quantum Monte Carlo techniques \cite{Sand} and the bosonization \cite{Gia}.
Effect of electron-electron interaction on two-dimensional finite clusters
of disordered spinless fermions has been studied by diagonalization in a truncated Hartree-Fock (HF) basis \cite{Sch}.
Persistent current
in disordered mesoscopic rings with spinless fermions, threaded by
a magnetic flux, has been calculated by using exact diagonalization 
technique for finite-sized systems \cite{Bouz1}. Similar systems
have been
studied by density matrix renormalization group (DMRG) \cite{Schmit} also.
 It seems interesting at this point to study the effect
of electronic interaction in a quasi-periodic lattice. This not only 
interpolates between the extreme cases of full grown order and random
disorder,
but also gives some insight into the electronic properties of
quasi-crystalline
superlattices which are now realizable in the laboratory.
Since the discovery of the quasicrystals \cite{Shet}, the electronic properties
of these novel systems have been studied with great interest. The low
temperature studies \cite{May} of the electronic conductivity of the
quasicrystalline
systems pose some interesting questions. On the other hand, recent fabrication
of quasicrystalline samples with local magnetic moments \cite{Sato} raise the
issue
of dealing with the electronic correlation in quasicrystalline materials.
Consequently, some works have already been carried out in this direction
which include bosonization and weak-coupling renormalization group (RG)
studies
of interacting spinless fermions with diagonal Fibonacci modulation
\cite{Vidal}. Hartree-Fock calculations for such Fibonacci-Hubbard chains have 
also been done \cite{Hira} to identify the nature of the single particle 
spectrum and that of the wavefunctions in case of small electronic correlation.
Recently, diagonal and off-diagonal Fibonacci modulations in an otherwise
periodic Hubbard model has been studied
 by a weak-coupling RG, complemented by a DMRG calculation \cite{Hida},
 which makes an attempt to identify gaps in the spin/charge sectors. However,
 to the best of our knowledge, a direct calculation of
the electronic conductivity for such a system is yet to be done. 

In view of this perspective we attempt a study of the competition between a
quasiperiodic ``disorder'' (e.g. Fibonacci modulation in site potentials) and
an on-site electronic correlation (e.g in the one band Hubbard model) in  one
dimension. The present study mainly concentrates on the calculation of the Drude
weight \cite{Kohn} within a generalized Hartree-Fock approximation (GHFA) for
studying the dc conductivity of such chains of {\it finite}
lengths at half-filling. We have also studied the
charge and spin density order parameters to characterize the ground state.
It may be noted here that the study of the Drude weight may not be meaningful 
for a very large system with strong correlation or disorder \cite{Gupta}. For
example, the strong on-site correlation could suppress the conductivity at
half filling for a large chain. Therefore, our calculation is restricted to 
small finite systems in order to resolve the issue of competition between
correlation and disorder. In fact previous Hartree-Fock studies of similar systems of large sizes \cite{Hira} did not attempt the calculation of 
conductivity and were restricted to the limit of small correlation only 
(neglecting spin polarization) which is not expected to fully capture the
essence of the competition between correlation and disorder.

In section II  we  define  the Hamiltonian and explain the calculations in
GHFA. Section III
describes the results obtained from GHFA for zero temperature
at half-filling. 
Section IV deals with the results at finite temperature.
 Section V summarizes the present work. 
\vskip .3in
\noindent
{\Large \bf II. The Model and the GHFA calculations}
\vskip .2in
\noindent
{\large \bf (i) Zero temperature calculation}
\vskip .1in
The Fibonacci sequence in any generation grows as:

\noindent
$A B A A B A B A A B A A B A B A A B A B A A B A A B A B A A B A A B..........~,$
growing with the generating rule where $A$ goes to $AB$ and  $B$ goes to $A$ in
the next generation.
For an infinite Fibonacci chain the ratio of $A$ to $B$ is $(\sqrt5+1)/2$, which
is known as the golden  mean.

We take a finite chain of $N$ sites, where $N$ is a Fibonacci number, in the 
form of a ring  that is threaded by a flux (in units of basic flux quantum
$\phi$ = $hc$/2$\pi e)$ \cite{Gupta}; the Hamiltonian looks like:
\begin{equation}
H=\sum_i \epsilon_i n_i + t\left(e^{i\phi} \sum_{i\sigma} c^{\dag}_{i \sigma} c_{{i+1} 
\sigma} + h.c. \right) + U \sum_i n_{i\uparrow} n_{i\downarrow},
\end{equation}
where $\epsilon_i$ is the site energy at the $i$-th site; it takes on the value $\epsilon_A$ or $\epsilon_B$
depending on whether it belongs to an $A$-type or a $B$-type site in the Fibonacci sequence.
The number operator $n_{i\sigma}$ =$c^{\dag}_{i\sigma} c_{i\sigma}$ and
$n_i = n_{i\uparrow} + n_{i\downarrow}$; $t$ is the hopping integral
between the nearest neighbour sites. The phase factor ($e^{i\phi}$)
 appears because of the flux threaded by the ring.
$U$ is the on-site Coulomb interaction.

On  employing the GHFA the Hamiltonian is decoupled into up-spin and down-spin
parts  with the modified site energies for the up- and the down-spin electrons. 
The modified site energies are given by:
\begin{eqnarray}
&&\epsilon'_{i\uparrow} = \epsilon_i + U <n_{i\downarrow}> \nonumber \\
&&\epsilon'_{i\downarrow} =\epsilon_i + U<n_{i\uparrow}> \nonumber
\end{eqnarray}

This decoupled Hamiltonian is then diagonalized (in a self-consistent manner) 
for both the
up and the down spin parts separately. We then calculate the ground state energy
$E_g$ by summing the states up to the Fermi level for both up and down spins
corresponding to a desired filling:
\begin{equation}
E_g = \sum^{\mu} E_{n\uparrow} + \sum^{\mu} E_{n\downarrow}-U \sum_i <n_ {i\uparrow}>
<n_{i \downarrow}>~.
\end{equation}
where $E_{n \sigma}$'s  are the single particle energy levels obtained by 
diagonalizing the decoupled Hamiltonian for the spin species $\sigma$.

The expressions for charge density wave (CDW)  and spin density wave (SDW)  
correlation functions are given by
\begin{eqnarray}
&& C(q)=\frac{1}{N} |\sum_{p,m}	e^{iq(R_p-R_m)} (n_p-1) (n_m-1)| \nonumber \\
&& S(q)=\frac{1}{N} |\sum_{p,m}	e^{iq(R_p-R_m)}(n_{p \uparrow}-n_{p \downarrow})(n_{m \uparrow}-n_{m \downarrow})| \nonumber \\
\end{eqnarray}
respectively, where $q$ is the wave-vector of the corresponding density wave.
 $R_p$ and $R_m$ are position vectors of the $p$-th and the $m$-th sites
 respectively.
The Drude weight which is a measure of the conductivity of the system
is given by \cite{Bouz1, Gupta, Scal}:
\begin{equation}
   D = -\frac{1}{N} \left(\frac{\partial^2 E_g}{\partial\phi^2}\right)_{\phi=0}
\end{equation}
\vskip .2in
\noindent
{\large \bf (ii) Finite temperature calculation}
\vskip .1in
	Upon diagonalization of the decoupled ``independent electron'' Hamiltonian one can calculate the partition function 
$$
Z=Tr [e^{-(\beta H -\mu N)}]~,
$$
where $\beta=1/kT$, $k$ being Boltzmann's constant and T, the temperature.
This leads to direct calculations of several thermodynamic quantities. The Drude
weight can be evaluated from
\begin{equation}
D= -\frac{1}{N} \sum_n \left[ \left(\frac{\partial ^2 E_{n \uparrow}}{\partial \phi ^2} \right)_{\phi=0}. \frac{1}{e^{\left(\frac{E_{n \uparrow} - \mu}{kT} \right)} +1}
+ \left( \frac{\partial ^2 E_{n \downarrow}}{\partial \phi ^2} \right)_{\phi=0}. \frac{1}{e^{\left(\frac{E_{n \downarrow} - \mu}{kT} \right)} +1} \right]
\end{equation}
\vskip .3in
\noindent
{\Large \bf III. Results for zero temperature}

Our results are based on calculations for different sizes of the Fibonacci
ring e.g. $N=$ 34, 55, 89, 233 and 377. 
We started our iteration with $\epsilon_A$ =0 and $\epsilon_B = \epsilon > 0$, 
with $U \ge 0$ the scale of energy being chosen by $t=1$.
After achieving the self-consistent solution we find out $n_{i\uparrow}$ and
 $n_{i \downarrow}$ for each site $i$, which enables us to calculate the CDW and SDW order parameters defined in (3). $E_g$ is obtained as a function of $\phi$,
by summing over states up to the Fermi level for both spins.
The double derivative of $E(\phi)$ with respect to $\phi$ at $\phi$=0 gives  the Drude
weight.
We have carried out the calculations for different system sizes with a
 fixed value 
of $\epsilon$  while varying $U$ and vice-versa.

It is interesting to observe how the competition between the correlation
 ($U$) and 
quasiperiodic modulation (${\epsilon}$) decides the conductivity of the
system (of a given size). It is also interesting to observe the dependence
of this effect on system size. 
 The
competition between disorder and correlation is expected to be most prominent
at half filling. This leads us to focus on this specific band-filling.
\vskip .1in
Fig. 1(a)  shows the variation of the Drude weight  with $(U/t)$ for different
values of
$\epsilon$ in a half filled chain of 34 sites.
In this plot one can observe
that the Drude weight is maximum at $U=0$ for $\epsilon=0$ and the
scattering due to the Fibonacci ``disorder'' (for $\epsilon \neq 0$) tends to
lower the conductivity compared to the case of $\epsilon=0$. 
However, in presence of the Fibonacci modulation in
the site potentials (for $\epsilon \neq 0$),
a competition between the correlation and the ``disorder'' enhances
 delocalization and,
thereby, the conductivity. As a result of this, there should be  a gradual rise in the Drude weight with increasing $U/t$. Such a 
competition between the Hubbard correlation and the  aperiodic 
Fibonacci modulation is indeed showing up in
the plots of the Drude weight in Fig. 1(a) with $\epsilon = 1.0$ and 1.5
 respectively. 
It can be argued on a heuristic ground that at $U \simeq 0$,
 there will be a considerable number of 
double occupancies (two electrons with opposite spins) that would favour to sit on
A-type sites with lower site-energies for $\epsilon \neq 0$; this tendency of ``pinning'' would 
decrease the Drude weight from that of the case with 
$\epsilon=0$. An increase in the value of $U$ would tend to melt the ``doublon''s and thereby undo the ``pinning'' effect. This would make the electrons
more and more delocalized resulting in a rise of the Drude weight with
increasing $U/t$ for $\epsilon \neq 0$.
After attaining a maximum value at a certain $U/t$, which depends on the
 value of $\epsilon$ (apart from the system size), the 
Drude weight will
again fall with increasing $U/t$ because of the usual suppression of the
conductivity by the Hubbard correlation. Eventually the Drude weight becomes
vanishingly small at a value of $U/t$, say $(U/t)_c$, which also
 depends on $\epsilon$
and system size.
The general features noted above for the case of a ring of 34 sites are also
observable in  larger rings of 55, 89 sites (Fig.s 1(b) and 1(c)). Even for
systems containing of as many as 233 or 377 sites (Fig.s 2(a) and 2(b)), we
could easily observe similar features i.e. the qualitative dependence of 
Drude weight on $\epsilon$ and $(U/t)$ is independent of system size.
 We must make a comment here
that the non-zero value of the Drude weight at half-filling is an effect of 
finite size. That is clearly indicated by the monotonic decrease in the
peak value of the Drude weight with system size for a given value of
$\epsilon$.
It is interesting to note here that the value of $(U/t)$ for which the 
Drude weight is a maximum decreases with increasing 
disorder strength ($\epsilon$). This feature, not clearly revealed for 
smaller system sizes can be prominently observed in Fig.s 2(a) and 2(b) where
we find a shift of the peak in the Drude weight towards lower values of $(U/t)$
with increasing $\epsilon$. This is so prominent in Fig. 2(b) that for large
enough $\epsilon$ we cannot recognize the gradual rise in the Drude weight.
The effect of disorder takes over to suppress the delocalization due to 
the competition between $U$ and $\epsilon$. For smaller system sizes
 such effect 
will take place for larger $\epsilon$. On the other hand the value of
 $(U/t)_c$, where the 
Drude weight  drops down to zero, depends on the system size. For a given 
value of $\epsilon$ it is observed that $(U/t)_c$ gradually decreases with
the increase of the system size (Fig. 3). So the competitive
effects of correlation and disorder show up in the conductivity of such systems of small sizes. The larger is the system size the weaker are the effects.
 Again for
 a ring of given length the competitive effect shows up for moderate
values of $\epsilon$ and $U$. A very high value of correlation monotonically
drives the system to an insulating phase; scattering due to strong disorder
also suppresses (even if $U$ is reasonably large) the interplay of correlation
and disorder (Fig. 2(b)). Such observations are in agreement with previous
studies \cite{Hida}. 
\vskip .1in
To
investigate the nature of the ground state further we have plotted the spin 
and charge structure factors $S(q)$ and $C(q)$ (defined in (3)) respectively.
In Fig.s 4(a) and 4(b) we have plotted $S(q)$ as a function of $q$ for different cases. In Fig. 4(a) we observe the sharp rise of the 
spin correlation function around $q$=$\pi$ for large $U$ (a value of $U$ when
the Drude weight is vanishingly small) for system sizes $N$=34, 89 and 233,
at $\epsilon=0$.
It is to be noted that precisely the point $q$=$\pi$ is missing
for finite odd sized systems. In the "clean" limit antiferromagnetic 
SDW phase should settle,
for larger $(U/t)$ which is revealed in the plot. We study the spin structure
factor for $\epsilon \neq$ 0 in Fig. 4(b) where we find smaller value of the
$q$=$\pi$ peak and some wiggles at $q$ $\neq \pi$ due to an imperfect 
antiferromagnetic modulation caused by aperiodic site potentials.
In presence of the 
Fibonacci modulation, there may arise a ``pinning'' tendency of the 
holes at special sites;  hence the reduction in the value of $S(q)$ around 
$q$=$\pi$ and the wiggles at $q$ $\neq \pi$. The effect is more pronounced
at higher values of $\epsilon$. Plots of $S(q=\pi )$ as a function of $(U/t)$
(Fig.s 5(a) and 5(b)) shows that the antiferromagnetic SDW grows after 
crossing a certain $(U/t)$ that depends on the system size and on the value
of $\epsilon$. These values $(U/t)$ match well with $(U/t)_c$ values obtained 
in the study of the Drude weight.
\vskip .1in
Fig.s 6(a), 6(b) and 6(c) show plots of the charge correlation function
$C(q)$  against $q$ for different system sizes and different
$U's$ at a reasonably large values of $\epsilon$. The quasiperiodic modulation
generates peaks of different heights at incommensurate values of $q$.
Density waves of such incommensurate (with lattice spacing) wave vectors
 get diminished in magnitude with increasing $U$. However the
 ${\it position}$  ${\it of}$  ${\it the}$
${\it peaks}$ do not change.
 With the increase of the system size the prominent peaks
do not disappear but get sharper and sharper. Moreover additional peaks of 
smaller magnitude appear to indicate excitation of new density waves for large
system sizes. This is expected because of hierarchical aperiodic structures of 
the Fibonacci generations.
\vskip .1in
In  explaining the decrease in the Drude weight with increasing ``disorder''
we conjectured the
 formation of ``pinned'' holes and ``doublons''. These ``pinning''
would follow the Fibonacci modulation and would be partly modified by the 
probability of hopping between adjacent sites of comparable site energies.
That this really takes place is evident from the plot of the density profile
against the site index (Fig. 7) for the 34-site ring. On the other hand
$A$ type sites favour retaining ``doublons''
 (with average occupancy greater than
 1). The
$AA$ pairs, on the contrary, show a lesser preference for the ``doublons'' to
facilitate nearest neighbour hopping which lowers the energy.
\vskip .2in
\noindent
{\Large \bf  IV. Finite temperature calculation of Drude weight}
\vskip .2in

	The Drude weight, as calculated from (5), is plotted against the 
temperature in Fig.s 8(a), 8(b), 9(a)  and 9(b) for system sizes $N=34$, 55,
89 and 233 respectively. Cases with zero and non-zero values
 of $\epsilon$ and $U$
are compared in the graphs. Since the effect of correlation is expected to
play an important role at the half-filling, the 34-site system is
studied at this special filling ($=\frac{17}{34}$) while for the 55, 89 and 233
-site
chains we concentrate on the fillings $n$= 28/55, 44/89 and 116/233
respectively which are the closest possible cases to (i.e. one particle away from) half-filling.
 
Fig. 8(a) shows the variation of the Drude weight against the temperature ($kT$)
for system size $N=34$. In absence of correlation ($U=0$) the Drude weight 
starts from its maximum value (i.e. $4t/\pi$) at half-filling for the periodic 
case ($\epsilon =0$). There is a gradual fall in the Drude weight with the rise in the temperature. This is because the thermal excitations tend to populate the
higher levels which diminish the current generated by the breaking of the time-reversal symmetry due to the external field.
 The rate of decrease of $D$, however,
is too small to be clearly detected in the graph. However, in the presence of
 the 
``Fibonacci disorder''($\epsilon \neq 0$) the scattering effect takes over and
this makes the conductivity fall in a much sharper fashion with the increase of
the temperature. On the other hand, at $T=0$, the Hubbard correlation introduces
a Mott 
gap \cite{Hubbard, Mott} at the middle of the band below which all levels (i.e. the lower band) are
occupied at half-filling.
 Therefore, the conductivity drops down to a much lower 
value (non-zero for a small finite system) in the case of $U \neq 0$ as compared 
to the non-interacting system. As the temperature increases, the excitations of 
the electrons take place across the Mott gap to populate the upper band. This 
results in an enhanced conductivity of the system. In the limit of very large 
temperature the Drude weight asymptotically approaches to that of 
the non-interacting case ($\epsilon = U =0$) from below. Simultaneous presence 
of aperiodic ``disorder'' and the Hubbard interaction leads to a competition 
that reduces the Mott gap. This is why the value of the Drude 
weight for $\epsilon \neq 0,~U \neq 0$ is slightly higher than that in the case 
of $\epsilon=0,~U \neq 0$. Now the Drude weight falls with the rise in 
temperature, the rate of fall being primarily dominated by the effect of
disorder.

	More or less similar pictures evolve for  other system sizes (Fig.s
8(b),9(a) and 9(b)). However, a few notable differences are there. Firstly, the 
effect of the  correlation becomes more pronounced with increasing system-size. This is why the Drude weight is vanishingly small for $U/t = 1.7$ ($> (U/t)_c$) for 
system sizes as large as $N=89$ and $233$ irrespective of the presence of disorder
(i.e. for both the cases of $\epsilon=0$ and $\epsilon \neq 0$). Secondly,
 the rate of fall 
of the Drude weight (with temperature) is enhanced for larger systems for which
the effect of disorder-induced scattering is stronger. It is noteworthy that in
Fig. 8(b) for $N=55$ the Drude weight initially falls and then increases with the rise in 
the temperature for the ordered case with large correlation
 ($\epsilon =0, U/t = 1.7$). This feature is in contradiction with the
 case depicted in Fig. 8(a) for $N=34$. This is because we have considered
 the precisely half-filled case
for $N=34$ while the case of $N=55$ corresponds to a case of one particle away
from the half-filling (filling = $28/55$). So, even in presence of the Mott gap
(comparable to the finite-size gaps) the system behaves like a metal at low temperatures. 
 Therefore, for the initial rise in temperature we observe a slow decrease in 
the Drude weight in Fig. 8(b). For higher temperatures the effect of thermal excitations of electrons across the Mott gap dominates. This results in the ultimate
rise in the Drude weight in this region. It is to be noted that this feature will gradually vanish with increasing system size for which the finite-size gaps
become smaller and smaller in comparison with the Mott gap. Indeed, this feature is merely detectable in Fig. 9(a) for $N=89$ while it completely disappears for
$N=233$ (Fig. 9(b)). The fact that the effect of the Mott gap becomes more dominant for larger systems (for the same value of the Hubbard parameter $U$) is also
evident from the case of $\epsilon = 0.5, U=1.7$ in Fig. 9(b) i.e. for $N=233$.
Here we find a small rise in the Drude weight with temperature even in presence
of disorder, a feature not observable in Fig.s 8(a), 8(b). For larger systems
the introduction of a small disorder essentially drives the system to become an
insulator which is can be sharply contrasted with that of a Mott insulator.
In the regime of higher temperatures we find that the conductivity of the 
disorder-driven insulator ($\epsilon=0.5~ U/t=1.7$) is less than that of the
correlation-driven (Mott) insulator ($\epsilon=0,~U/t=1.7$) by several orders of magnitude. This is, indeed, a very good illustration of differentiating the
two types of insulators$-$ namely, disorder-driven insulator and Mott insulator.
\vskip .3in
\noindent
{\Large \bf  V. Conclusion}
\vskip .2in

Summarizing, we have studied finite-sized Fibonacci-Hubbard chains at 
 half filling (or one particle away from half-filling). In the present work,
 within the mean field approach of GHFA,
we calculate the ground state properties e.g. 
 the spin and
charge correlation functions, and the dc conductivity both at zero and finite 
temperatures.
We find that the dc conductivity, as obtained from the
Drude weight \cite{Bouz1, Gupta}, shows a strong dependence on the 
 competition between the Fibonacci modulation
and the Hubbard correlation. This tends to enhance the conductivity at zero 
temperature in the regime of intermediate coupling. However, in the asymptotic 
cases of 
very weak and strong Hubbard correlations (in presence of a reasonable
modulation in the site potentials), the conductivity is reduced due to
diagonal aperiodicity (driving towards a Fibonacci modulated charge ordering) 
and electronic correlation (bringing in antiferromagnetic fluctuations) 
respectively \cite{Hida}. The value of $(U/t)_c$, where the conductivity
of the system drops down to zero, decreases with the system size. This suggests
 that the true effect of competition of the correlation and disorder could
be appreciably observed in systems of smaller size
 (typically$\sim$100 sites).
The SDW correlation strongly grows with Hubbard correlation at $q$=$\pi$ 
at half-filling. This indicates the dominance of antiferromagnetic 
instabilities at large $U$ values.
The charge correlation $C(q)$ shows spikes characteristic of Fibonacci
modulation. This charge modulation although reduced by the presence of
 finite $U$ does not go off even at a value of $U$ for which anti-ferro SDW
starts dominating. There appear more and more characteristic spikes
 in $C(q)$ as the
system size increases. However, the  spikes observed for smaller system sizes
never disappear.
This suggests that the system captures the self similar pattern of the
Fibonacci hierarchy and this is not masked by correlation.
At finite temperatures, the Drude weight slowly decreases with rise in the 
temperature, in the periodic limit in absence of correlation. The presence
of Hubbard correlation opens up a Mott gap which quenches the conductivity
at low temperatures.
Therefore, with the increase of temperature, there are excitations of electrons
across the Mott gap to give rise to an initial increase in the conductivity.
This semiconductor-like behaviour is in sharp contrast to the periodic case. On the other hand,
 the scattering induced by the Fibonacci disorder makes the picture completely
different. In presence of Fibonacci modulation alone the Drude weight sharply
decreases with temperature. This feature shows up even in presence of the
Hubbard
 interaction for smaller system sizes. However, for larger systems at very
low temperatures (in presence of correlation) the Mott gap 
reduces the
 conductivity to such a small value that the disorder-driven fall in the 
Drude weight is almost invisible. In the high temperature region the behaviour
of the Mott insulator and that of the disorder-driven insulator are in sharp 
contrast to each other. The results obtained for a 55-site system 
(studied for one particle away from half filling) indicate that the effect of 
variation in the band filling could give rise to some notable features.	As
 revealed in the present study, the competition between electronic
correlation and diagonal aperiodicity leads  to interesting effects
in the properties of such systems at both $T=0$ and $T \neq 0$ for
smaller system sizes.
It is, therefore, worthwhile to make further studies of such systems. For 
example, it would be interesting to see the effect of different types of 
quasiperiodic modulations and the effect of external fields in such systems,
using similar mean-field approaches as well as techniques beyond the mean-field
approximations \cite{Sand, Hida}. Also a detailed study of the effect of 
band filling away from the half filled case would be an interesting
one at finite temperatures.

\newpage

\noindent
{\Large \bf {Figure Captions}}
\vskip .2in

\noindent

\noindent
Fig. 1. Plot of the Drude weight $D$ (at zero temperature) vs. $U/t$ for different values of $\epsilon$ 
for a Fibonacci-Hubbard ring of (a) 34 sites (with 34 particles),
(b) 55 sites (with 56 particles) and (c) 89 sites (with 88 particles). 
A finite value of $D$ for low $U/t$ at half-filling
is a finite-size effect. For non-zero
values of $\epsilon$ the competition between Hubbard correlation and the diagonal
Fibonacci modulation is clearly seen from low to intermediate values of $U/t$ 
$-$ upto a value $(U/t)_c$.
\\

\noindent
Fig. 2. Plot of the Drude weight $D$ (at zero temperature)  vs. $U/t$ for the
Fibonacci-Hubbard chain of (a) 233 sites (with 232 particles) and (b) 377 sites (with 376 particles) for different values of $\epsilon$.  
The shift of the peak in the Drude weight towards $U/t=0$ is clearly visible for
higher values of $\epsilon$.
\\

\noindent
Fig. 3. Plot of $(U/t)_c$ vs. $N$, the system size (at/ one particle away from 
half-filling).
\\

\noindent
Fig. 4. Plot of the SDW correlation function $S(q)$ (at zero temperature) as a 
function of the 
wave-vector $q$ for different systems sizes (a) in absence of disorder ($\epsilon = 0$) and (b) in presence of Fibonacci modulation ($\epsilon =1.5$). Value of
the Hubbard correlation $U =3.0$ (the scale of energy is given by $t=1.0$). The
peak at $q = \pi$ shows antiferromagnetic spin modulation induced by large $U$.
This correlation is reduced by the presence of Fibonacci ``disorder'' and also
there appear wiggles at $q \neq \pi $. Such wiggles are merely visible in (b).
\\

\noindent
Fig. 5. Plot of $S(q= \pi )$  (at zero temperature)  vs. $U/t$ for different
values of $\epsilon $
for the Fibonacci-Hubbard chain of (a) 34 sites,
and (b) 89 sites. For the
Hubbard model ($\epsilon=0$) $S(q = \pi )$ grows with $U/t$  at half-filling.
The  sharp rise in $S(q = \pi )$ follows after a certain value of $(U/t)$
which is in close agreement with the $(U/t)_c$ observed in Fig. 1. 
\\

\noindent
Fig. 6. Plot of the CDW correlation function $C(q)$ (at zero temperature) vs. the wave-vector $q$ for the Fibonacci-Hubbard chain of (a) 34 sites, (b) 89 sites, and (c) 233
sites. There are characteristic peaks for a large value of the ``disorder''
($\epsilon = 1.5$) which do not disappear with the increase in the correlation
parameter $U$. 
\\

\noindent
Fig.7. Plot of the particle density $n_i$ against the site index $i$ for the 
Fibonacci-Hubbard chain of 34 sites at zero temperature at half-filling with
 $\epsilon=1.0$ and $U=3.0$ (in unit of $t$). The density profile shows formation of a charge density modulation which closely resembles the Fibonacci pattern.
\\

\noindent
Fig. 8. Plot of the Drude weight $D$  vs. the temperature
$T$ (scaled up by the Boltzmann constant $k$) for the Fibonacci-Hubbard chain
of  (a) 34 sites (at half-filling),  and (b) 55 sites (one particle away from 
half-filling) for different choices of $\epsilon$ and $U$ (scale of energy $t=1.0$).
\\

\noindent
Fig. 9. Plot of the Drude weight $D$  vs. the temperature
$T$ (scaled up by the Boltzmann constant $k$) for the Fibonacci-Hubbard chain
of  (a) 89 sites,  and (b) 233 sites (for one particle away from 
half-filling for both the cases) for different choices of $\epsilon$ and $U$ (scale of energy $t=1.0$).
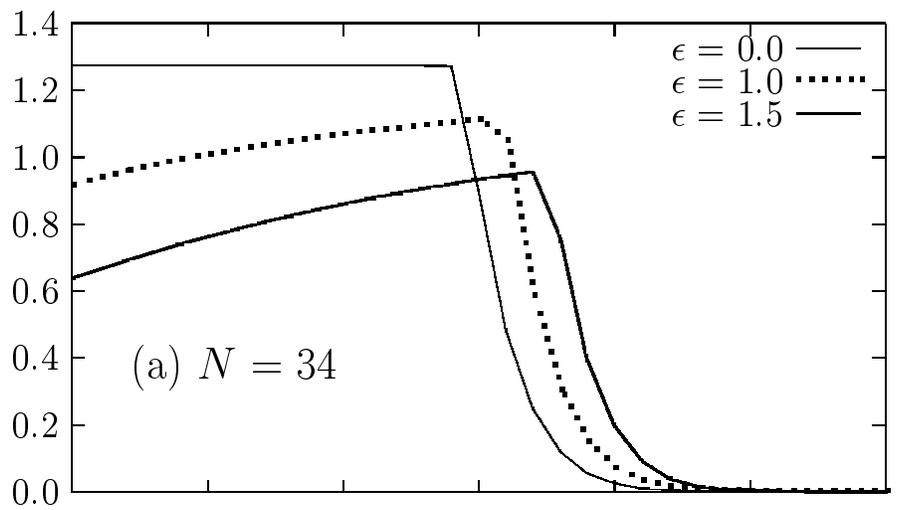
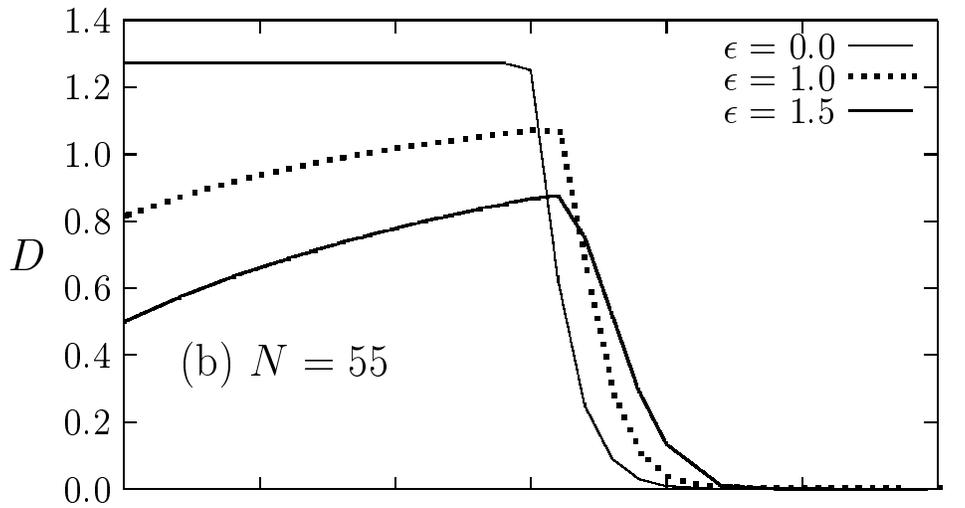
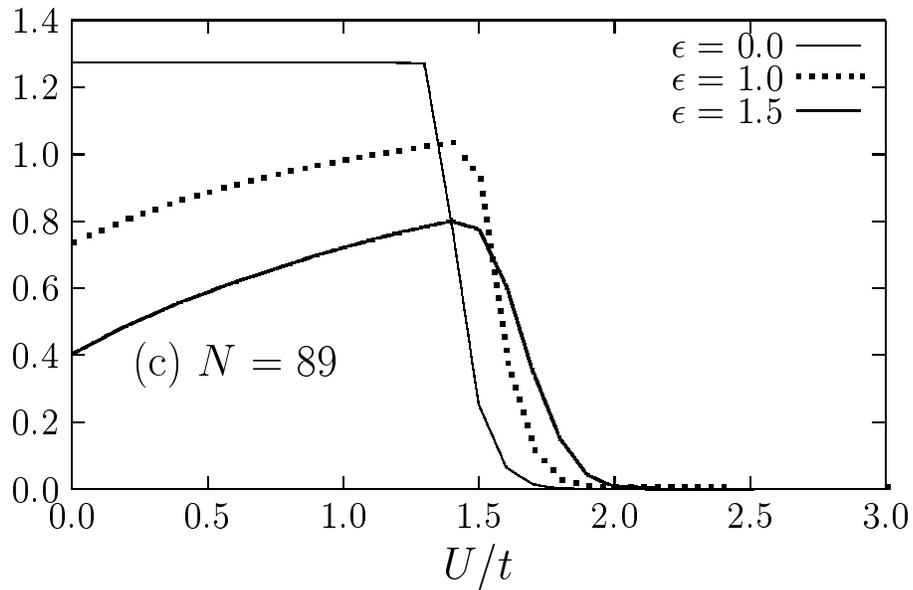
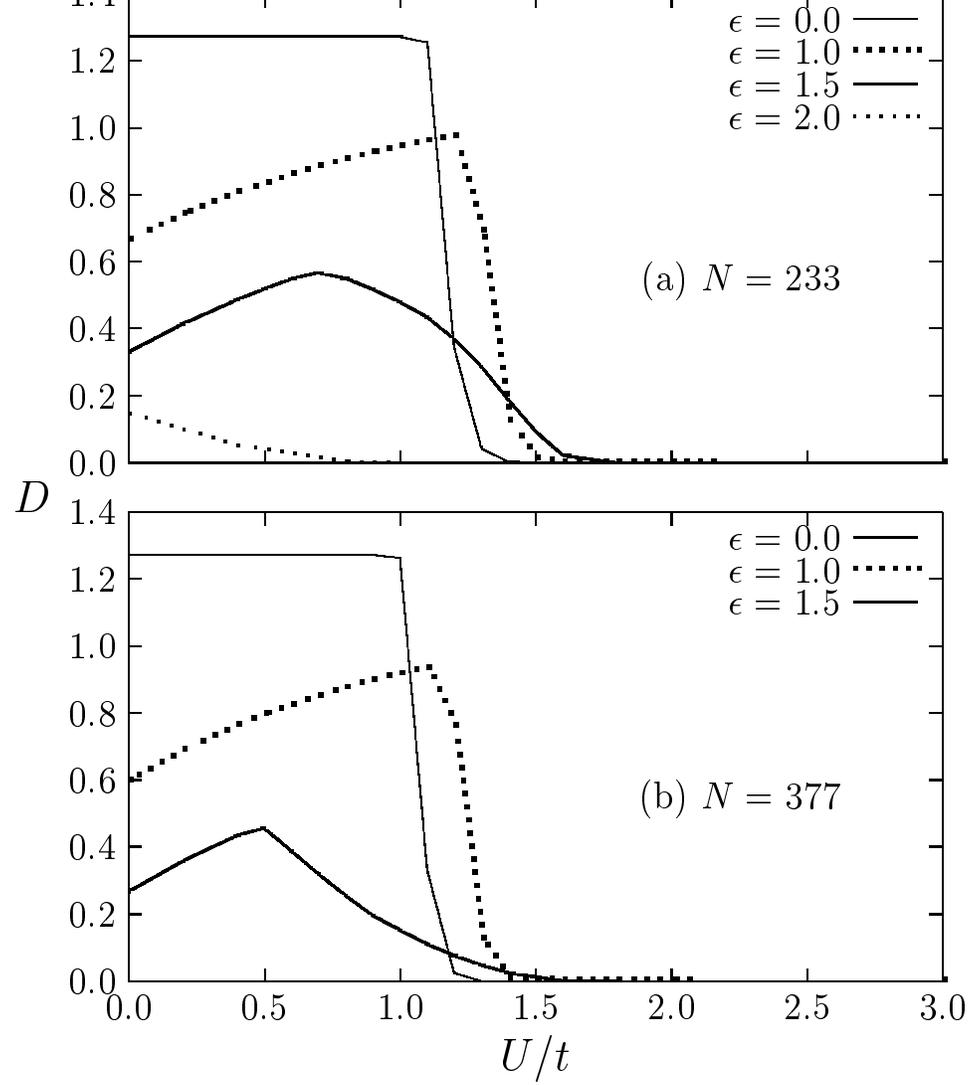
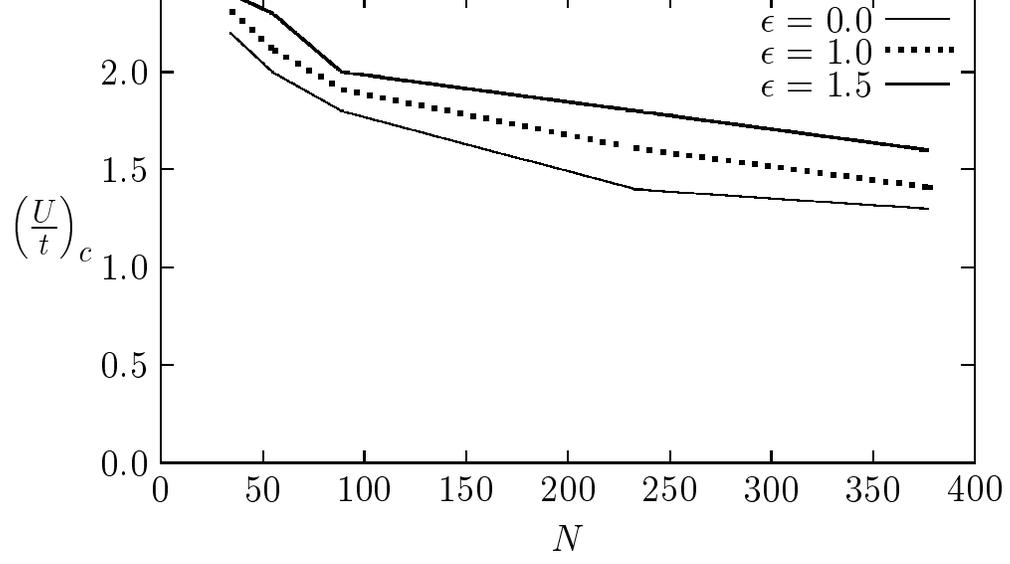
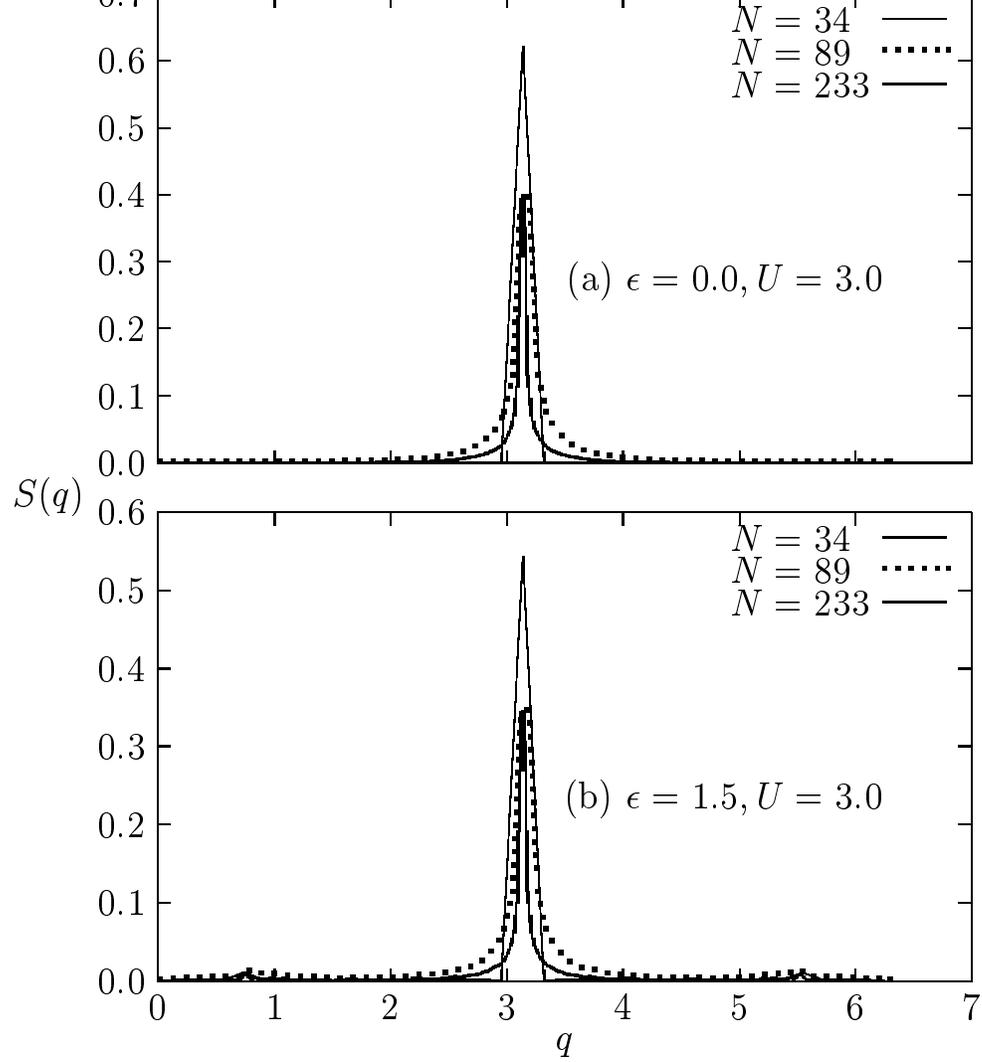
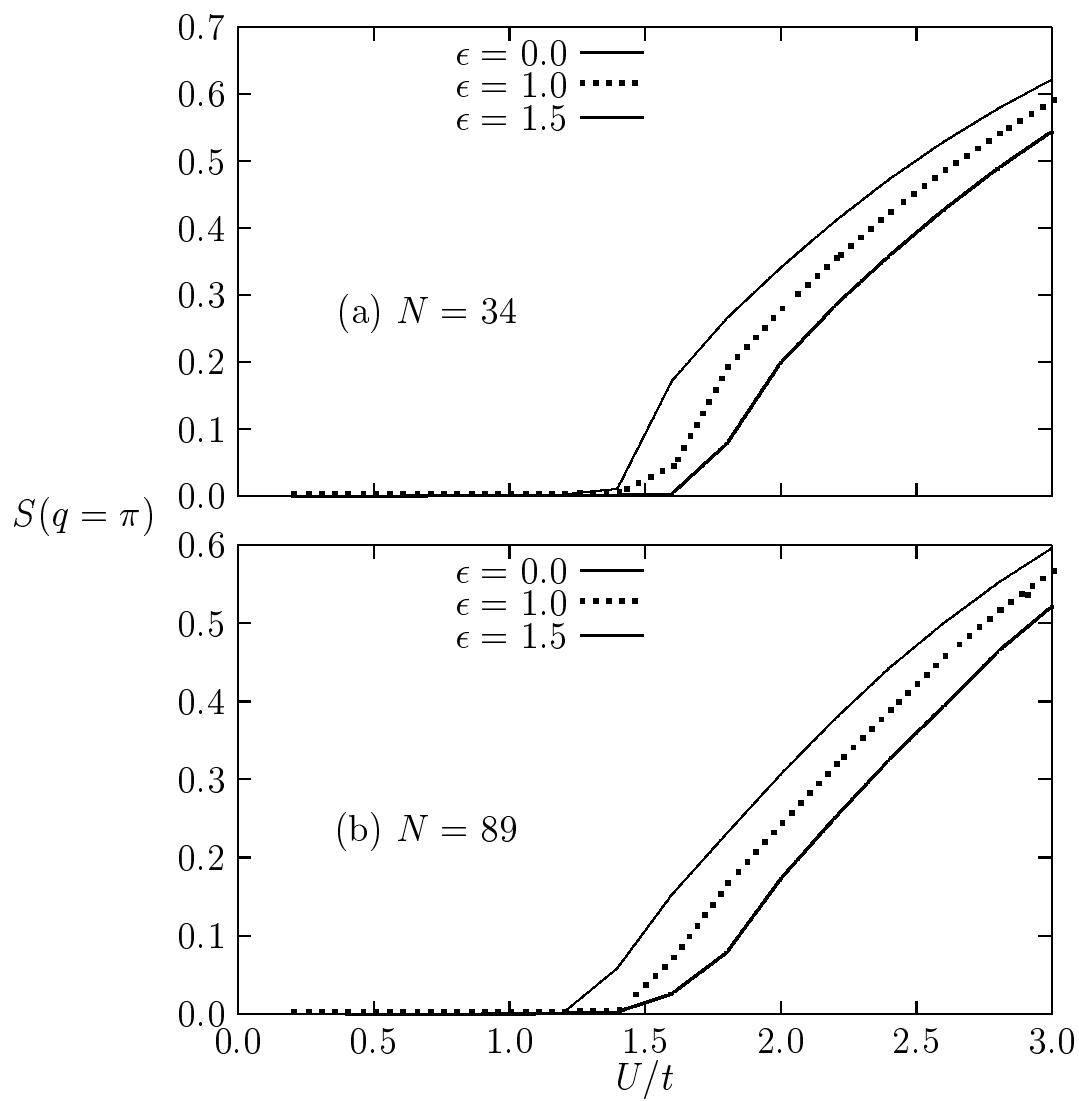
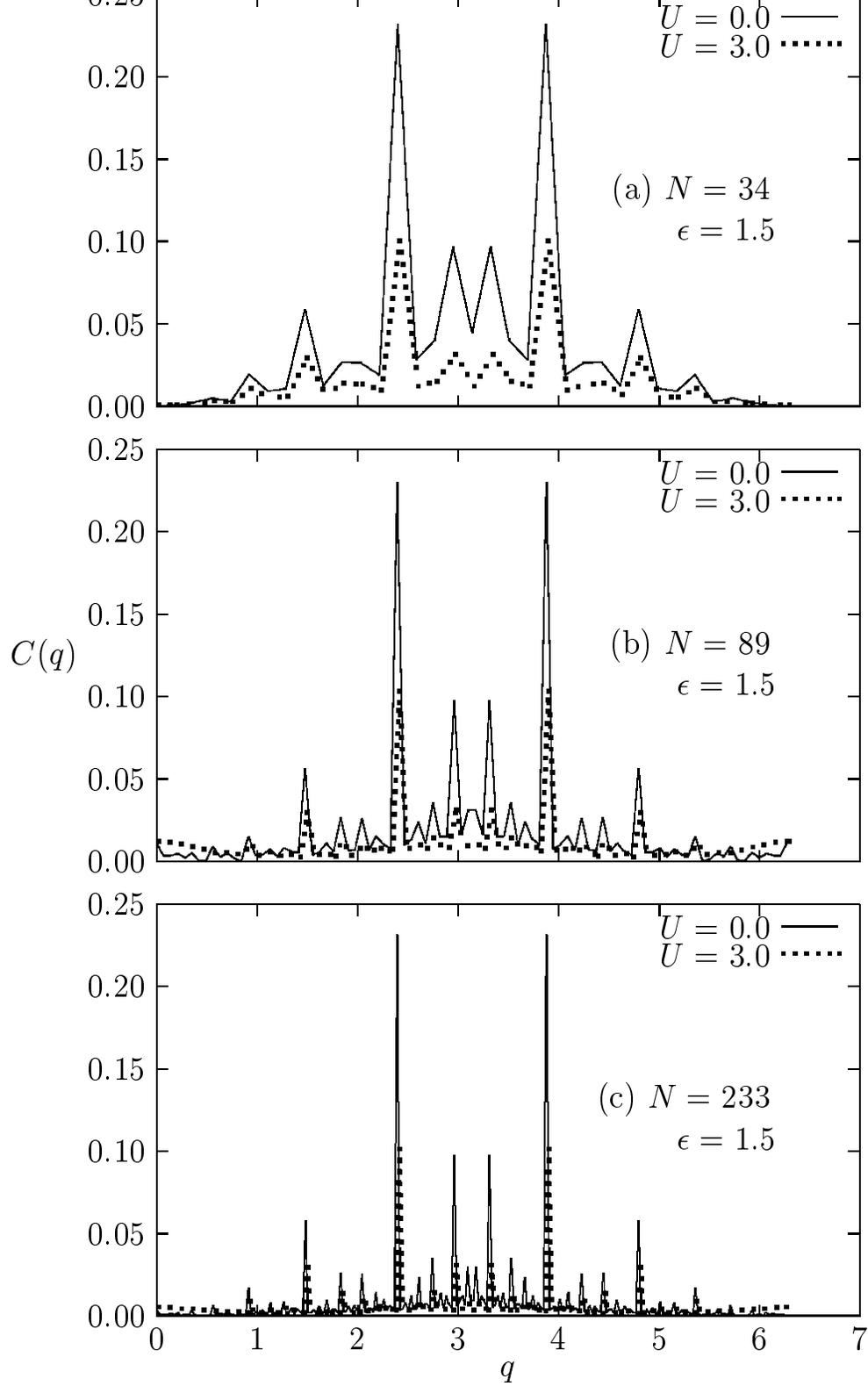
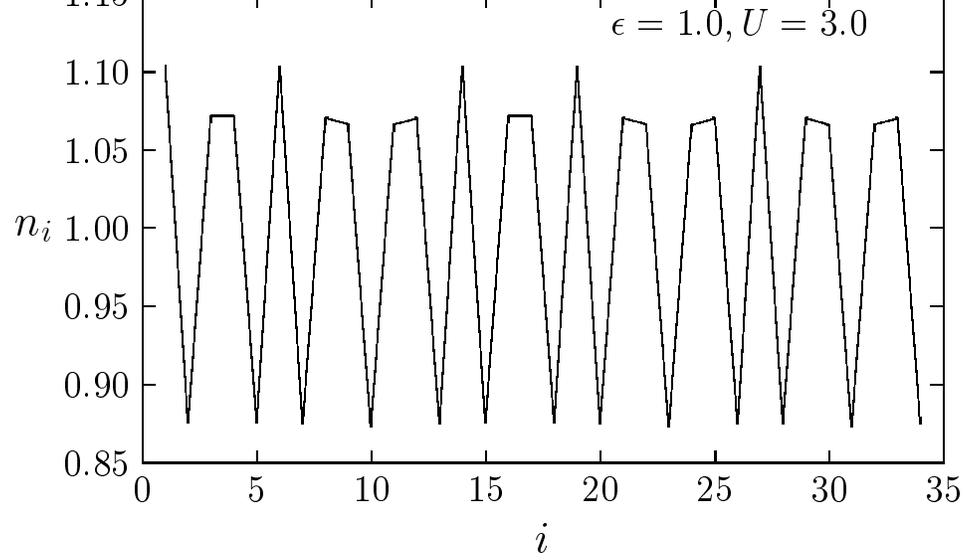
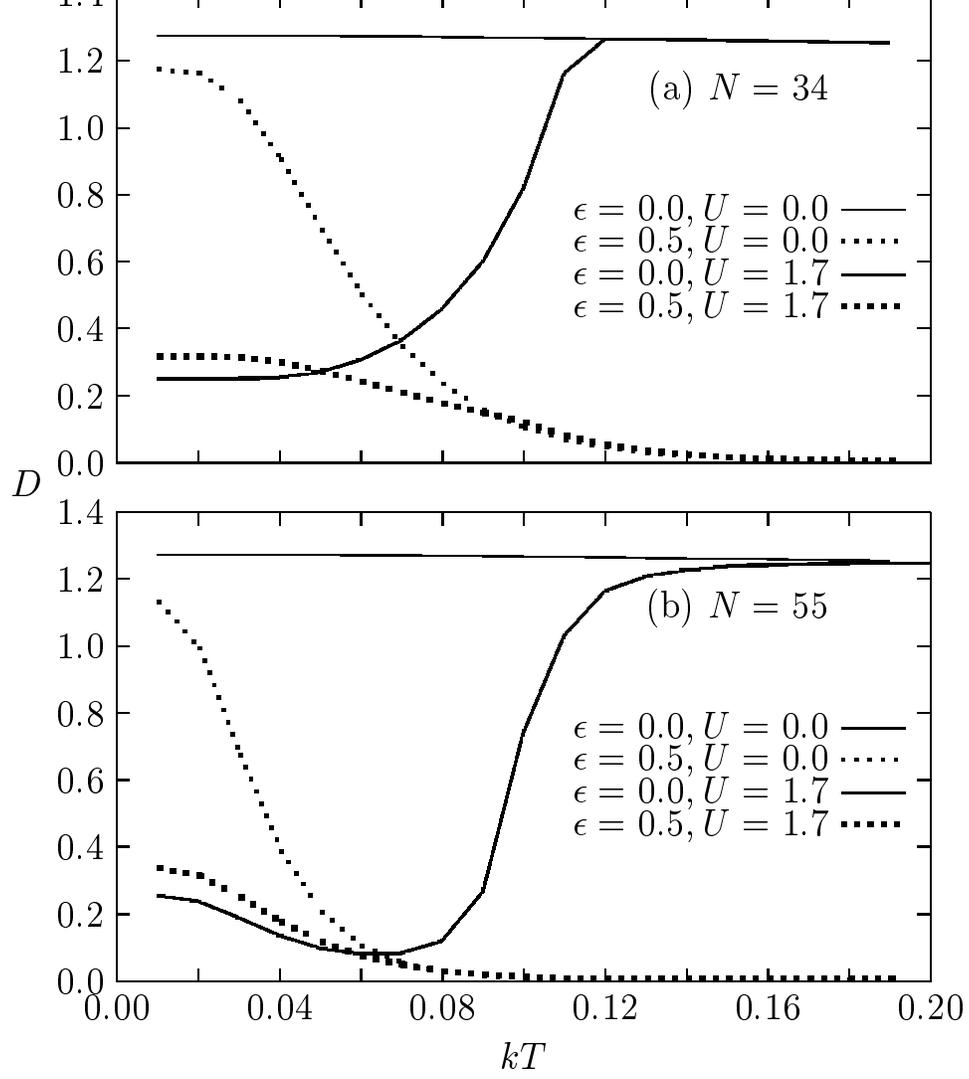
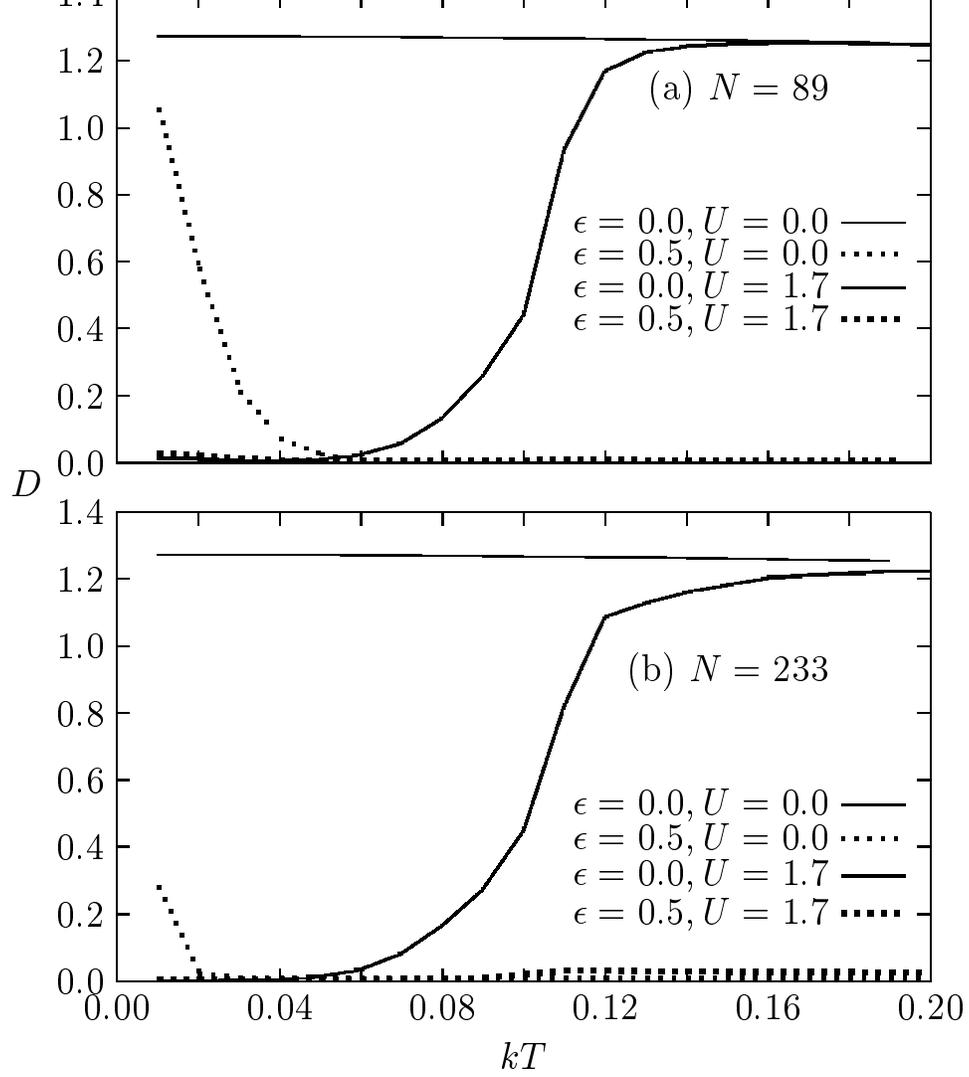
\end{document}